\documentclass[preprint,showpacs,preprintnumbers,amsmath,amssymb]{revtex4}

\usepackage{graphicx}
\usepackage{bm}

\begin{document}

\title{Temporal precision of spike response to fluctuating input in
pulse-coupled networks of oscillating neurons}
\author{Jun-nosuke Teramae}
\email{teramae@riken.jp}
\author{Tomoki Fukai}
\affiliation{Brain Science Institute, RIKEN, Saitama, Japan}

\begin{abstract}
 A single neuron is known to generate almost identical spike trains when
 the same fluctuating input is repeatedly applied. Here, we study the
 reliability of spike firing in a pulse-coupled network of oscillator
 neurons receiving fluctuating inputs. We can study the precise
 responses of the network as synchronization between uncoupled copies of
 the network by a common noisy input. To study the noise-induced
 synchronization between networks, we derive a self-consistent equation
 for the distribution of spike-time differences between the
 networks. Solving this equation, we elucidate how the spike precision
 changes as a function of the coupling strength.
\end{abstract}



\maketitle
 
Reliable information processing requires a code that can be represented
and transmitted reliably within the precision of the devices. In the
brain, single neurons can generate highly precise spike trains when they
are repeatedly activated by the same fluctuating input\cite{bryant}. Neurons,
however, work collectively rather than individually in their network. It
remains unclear whether precise elements put in a network can still
respond precisely since mutual couplings may affect the response of the
individual neurons\cite{tiesinga08}. To answer this question, we investigate the
temporal precision of responses of a pulse-coupled network of
oscillators when a set of independent fluctuating inputs, i.e., frozen
noise, is repeatedly applied to the network. To study this problem
analytically, we introduce uncoupled copies of the network that commonly
receive the set of fluctuating inputs. Suppose that the original network
repeatedly generates identical precisely-timed spike responses. Then, we
can interpret the collection of these responses across trials as
responses of the individual copies in a trial. Thus, in-phase
synchronization between identical networks implies precisely-timed
responses of a single network across trials.

Such noise-induced synchronization was previously studied between single
oscillators\cite{teramae,piko,nakao}. Here, we study the noise-induced
synchronization between networks of oscillators to clarify whether each
network is able to encode information about fluctuating inputs into
precisely-timed spikes. Noise-induced synchronization can appear in
lasers\cite{laser}, chemical reactions\cite{chemical}, gene
networks\cite{gene}, electronic circuits and neural
systems. For instance, noise-induced
synchronization of neural oscillators may play an active role in
olfactory information processing\cite{galan}. An attempt was made to
employ such an oscillating synchronization for a dynamic clock in
digital VLSI devices\cite{utagawa}. Thus, clarifying the underlying
mechanism has significant implications for a variety of dynamical systems.

We develop a mean-field theory of noise-induced synchronization between
the copies of the pulse-coupled oscillator network. We can analytically
derive a self-consistent equation for the distribution of phase
differences between the corresponding oscillators and obtain the
distribution as a function of the coupling strength. In so doing, we
assume that the average of the connections projecting to each oscillator
vanishes. The distribution allows us to reveal the nontrivial effect of
mutual couplings on the noise-induced synchronization of oscillator
networks or, equivalently, the temporal precision of responses in an
oscillator network.

Let us consider multiple trials in which a pulse-coupled network of
homogeneous $N$
oscillating neurons receive the same fluctuating input in all trials:
\begin{equation}
 \dot{\bm{x}}_i^{\left(\alpha\right)}
  =\bm{F}\left(\bm{x}_i^{\left(\alpha\right)}\right)
    +\sum_{j=1}^N g_{ij}
    \sum_n \delta\left(
		    t-t_{j,n}^{\left(\alpha\right)}
		   \right)
    +\xi_i\left(t\right),
 \label{eq.1}
\end{equation}
where $i=1,\cdots ,N$, $\bm{x}_i^{\left(\alpha\right)}$ is the state
variable of the $i$th neuron in the $\alpha$th trial,
$\bm{F}\left(\bm{x}_i^{\left(\alpha\right)}\right)$is the intrinsic
dynamics of the neuron, $g_{ij}$ is the coupling strength from 
the $j$th to the $i$th neuron, $t_{j,n}^{\left(\alpha\right)}$ is
the $n$th spike time of the $j$th 
neuron in the $\alpha$th trial, and $\xi_i\left(t\right)$ is the fluctuating input to the $i$th
neuron. $\bm{F}\left(\bm{x}\right)$ has a stable limit-cycle solution
$\bm{x}_0\left(t\right)$ satisfying
$\dot{\bm{x}_0}=\bm{F}\left(\bm{x}_0\right)$ with period 
$T$. Isolated neurons thus fire regularly with a firing rate $r=1/T$. We use
the zero mean white Gaussian noise with variance $D$ as the fluctuating
input $\xi_i\left(t\right)$, which is independent among neurons whereas it should be the same
across trials, $\left<\xi_i(t)\right>=0$,
$\left<\xi_i(t)\xi_j(s)\right>=D\delta_{ij}\delta(t-s)$. Since Eq. (\ref{eq.1})
 is a stochastic differential equation,
we have to clarify its interpretation. We use the Stratonovich
interpretation, namely, we define the white noise as the limit of
colored noise with infinitesimal correlation time\cite{strato}. Components of the
zero mean matrix $g_{ij}$ take both positive and negative values independently,
while they should also be exactly the same across trials. We denote the
variance of the matrix as $G$, $\left<g_{ij}\right>=0$,
$\left<g_{ij} g_{kl}\right>=G\delta_{ik}\delta_{jl}$. Even though the
same input and the same network are shared by all trials, it is unclear
whether spike times would coincide across trials because initial values
are different among trials. 

In order to obtain a unified description of the problem, we apply the
standard phase reduction method\cite{kuramoto} to Eq. (\ref{eq.1}). Regarding both the
fluctuating inputs and the interactions as perturbation $p_i^{\left(\alpha\right)}$ to the
limit-cycle oscillators,
we obtain the following stochastic differential 
equations for phase variables:
\begin{eqnarray}
 \dot{\phi}_i^{\left(\alpha\right)}
  &=&1+Z\left(\phi_i^{\left(\alpha\right)}\right)
  p_i^{\left(\alpha\right)}\left(t\right),\label{eq.2}\\
 p_i^{\left(\alpha\right)}\left(t\right)
  &=&\sum_{j=1}^N
  g_{ij}\sum_n\delta\left(t-t_{j,n}^{\left(\alpha\right)}\right)
  +\xi_i\left(t\right)\nonumber
\end{eqnarray}
where the phase $\phi_i^{\left(\alpha\right)}$ is defined to increase by
$T$ for every cycle of $\bm{x}_i^{\left(\alpha\right)}$ around 
the limit cycle. Natural angular velocity is thus equal to 1 and the
spike time $t_{j,n}^{\left(\alpha\right)}$ satisfies the relation
$\phi_i^{\left(\alpha\right)}\left(t_{i,n}^{\left(\alpha\right)}\right)=nT$.
The phase response function, or the phase sensitivity,
$Z(\phi)=\nabla_{\bm{x}}\phi|_{\bm{x}=\bm{x}_0(\phi)}$,
quantifies the phase response to
perturbations\cite{kuramoto}. 

Spike time difference across trials is quantified as the distribution
function of phase differences across trials in the phase description. If
corresponding oscillators in different trials synchronize with each
other in phase, the distribution will be the delta-function and spike
trains will be the same across trials. If, in contrast, oscillators
across trials do not synchronize perfectly, the distribution will have a
finite width, which characterizes variation of spike times across
trials. 
The synchronization across trials is an extension of common-noise
induced synchronization between uncoupled oscillators.
It is noteworthy that the synchronization mentioned here is
different from widely studied synchronization within a network; rather
this is synchronization across trials that are mutually uncoupled by
definition. Actually, oscillators within the network tend to
desynchronize each other because of independent fluctuating inputs they
received.

To obtain the distribution function of phase differences, we first
derive the Fokker-Planck equation satisfied by the distribution
\cite{kuramoto_un,strato}. Without 
loss of generality, we focus on phase differences between two trials,
$\psi_i=\phi_i^{(2)}-\phi_i^{(1)}$. The Fokker-Planck equation for the
distribution $P\left(\psi_i,t\right)$ is expressed as
$\frac{\partial P}{\partial t}=-\frac{\partial}{\partial
\psi_i}\left(A_1 P\right)+\frac{\partial^2}{\partial\psi_i^2}\left(A_2
P\right)$,
where the 1st and the 2nd Kramers-Moyal coefficient, $A_1$ and $A_2$,
are defined as moments of $\psi_i$ increment in an infinitesimal time step,
$A_n\left(\psi_i\right)=\lim_{t\to 0}\frac{1}{n!t}\left<\left(\psi_i\left(t\right)-\psi_i\left(0\right)\right)^n\right>|_{\psi_i\left(0\right)=\psi_i}$.
We now invoke the averaging assumption to calculate the
coefficients\cite{kuramoto_un}\cite{kuramoto,nakao,ermentrout07}. With
sufficiently small perturbations, the distribution 
function $P\left(\psi_i,t\right)$ varies slowly
compared with the oscillator natural period, $T(=1/r)$. We can thus
replace the infinitesimal time-step increment in $A_n$ by the average
increment in the period:
\begin{equation}
 A_n\left(\psi_i\right)
  \simeq\frac{1}{n!T}
  \left<\left(\psi_i\left(T\right)-\psi_i\left(0\right)\right)^n\right>|
  _{\psi_i\left(0\right)=\psi_i}.
 \label{eq.3}
\end{equation}
The increment is calculated from Eq. (\ref{eq.2}) as follows. Integration of
Eq. (\ref{eq.2}) from the initial phase
$\phi_i^{\left(\alpha\right)}\left(0\right)$ gives
\begin{equation}
 \phi_i^{\left(\alpha\right)}\left(t\right)
  =\phi_i^{\left(\alpha\right)}\left(0\right)
  +t
  +\int_0^t dx Z\left(\phi_i^{\left(\alpha\right)}\left(x\right)\right)
  p_i^{\left(\alpha\right)}\left(x\right).
 \label{eq.4}
\end{equation}
To evaluate $\phi_i^{\left(\alpha\right)}\left(x\right)$ which appears
in the integrand, we substitute Eq. (\ref{eq.4}) recursively into the
right hand side and obtain
\begin{equation}
 \phi_i^{\left(\alpha\right)}\left(t\right)
  =\phi_i^{\left(\alpha\right)}\left(0\right)
  +t
  +\int_0^t dx Z\left(\phi_i^{\left(\alpha\right)}\left(0\right)+x\right)
  p_i^{\left(\alpha\right)}\left(x\right)
\label{eq.5}
\end{equation}
up to the 1st order of $p_i^{\left(\alpha\right)}$
\cite{kuramoto_un}.
Using the assumption $\left<g_{ij}\right>=0$, we find
\begin{align}
&<p_{i}^{\left(1\right)}\left(x\right)p_{i}^{\left(2\right)}\left(y\right)>\nonumber\\
&=\left<\xi_i\left(x\right)\xi_j\left(y\right)\right>
+\left<\sum_{j,n,k,m}g_{ij}g_{ik}
\delta\left(x-t_{j,n}^{\left(1\right)}\right)
\delta\left(y-t_{k,m}^{\left(2\right)}\right)\right>\nonumber\\
&=D\delta\left(x-y\right)
+\left<\sum_{j,n,m}g_{ij}^2
\delta\left(x-t_{j,n}^{\left(1\right)}\right)
\delta\left(y-t_{j,m}^{\left(2\right)}\right)\right>\nonumber\\
&=D\delta\left(x-y\right)
+GNrS\left(x-y\right),\nonumber
\end{align}
where the normalized cross-correlation function between spike
trains of trials is defined as
$S\left(x-y\right)=\left<\sum_{j,n,m}\delta(x-t_{j,n}^{\left(1\right)})\delta(y-t_{j,m}^{\left(2\right)})/(Nr)\right>$,
which yields the distribution of spike time difference between
trials.
Substituting both $\psi_i=\phi_i^{(2)}-\phi_i^{(1)}$ and Eq. (\ref{eq.5})
into Eq. (\ref{eq.3}) and
using the above equation, we obtaine coefficients explicitly as
$A_1\left(\psi\right)=0$ and
\begin{align}
 A_2\left(\psi\right)
  &=\frac{D+NGr}{T}\int_0^T dx Z\left(x\right)^2
 -\frac{D}{T}\int_0^T dx
  Z\left(x\right)Z\left(x+\psi\right)\nonumber\\
 &-\frac{NGr}{T}\int_0^T dx\int_0^T dy
  Z\left(x\right)S\left(x-y\right)Z\left(y+\psi\right).\label{eq.6}
\end{align}
From Eq. (\ref{eq.6}), we obtain
the explicit form of the Fokker-Planck equation. The distribution $P$ is given as
the stationary solution of the equation,
\begin{equation}
 P\left(\psi\right)
  =c/A_2\left(\psi\right),
 \label{eq.7}
\end{equation}
where $c$ is a normalization constant.

Because $A_2\left(\psi\right)$ still includes the unknown function $S$,
Eq. (\ref{eq.7}) is not a closed form of the distribution $P$. However, we can
derive a quite simple relationship between $P$ and $S$. From the
relationship
$\phi_i^{\left(\alpha\right)}\left(t_{i,n}^{\left(\alpha\right)}\right)=nT$,
we obtain
$\phi_i^{\left(2\right)}\left(t_{i,n}^{\left(2\right)}\right)-\phi_i^{\left(1\right)}\left(t_{i,n}^{\left(1\right)}\right)=0$.
Expanding this with respect to spike time difference when the difference
is small and using the fact that the natural velocity of phase is unity,
we obtain
$t_{i,n}^{\left(1\right)}-t_{i,n}^{\left(2\right)}=\phi_i^{\left(2\right)}-\phi_i^{\left(1\right)}=\psi_i$,
which means that the spike time difference is equal to the phase
difference. Therefore their respective distributions should
also be the same;
\begin{equation}
 P=S.
 \label{eq.8}
\end{equation}
Using Eq. (\ref{eq.8}) is a type of the
mean-field approximation which has been widely used in physics and
theoretical neuroscience\cite{amit_brunel}. Unlike the previous studies, however, we do
not treat statistics only within the network; rather we can treat
statistics between uncoupled trials.

Equations (6-8) give a self-consistent equation for the
distribution function $P$, and for $S$ simultaneously. Since $P$ is a
$T$-periodic function, we can use the Fourier expansion to simplify the
equation. The $n$th Fourier component $f_n$ of a $T$-periodic function $f$
is defined from $f\left(x\right)=\sum_{n=-\infty}^{\infty}f_n e^{i2\pi nx/T}$.
Noting that $P$ is an even function due to symmetry between
trials, we can obtain the expression $P$ from Eq. (\ref{eq.6}) and (\ref{eq.7}) as
\begin{equation}
 P\left(\psi\right)
  =\frac{\tilde{c}}{\sum_{n=1}^{\infty}\left|Z_n\right|^2
  \left(1-W_n\cos\left(\frac{2\pi n}{T}\psi\right)\right)},
 \label{eq.9}
\end{equation}
where $\tilde{c}$ is a
normalization constant and $W_n=(1+qP_n)/(1+qr)$.
The variable $q$ is defined as the ratio of variance between
internal input and external input per neuron $q=NG/D$, which acts as the control
parameter of our problem. We then obtain the
self-consistent equation for the Fourier component of $P$ as the final form: 
\begin{equation}
 P_m
  =\frac{\int_0^{2\pi}
  \frac
  {\cos m\psi d\psi}
  {\sum_{n=1}^{\infty}\left|Z_n\right|^2\left(1-W_n\cos n\psi\right)}}
  {\int_0^{2\pi}
  \frac
  {T d\psi}
  {\sum_{n=1}^{\infty}\left|Z_n\right|^2\left(1-W_n\cos n\psi\right)}}
  =H_m\left(\{P_n\}\right).
 \label{eq.10}
\end{equation}
The infinite Fourier series which appears in
 Eq. (\ref{eq.9}) and (\ref{eq.10}) is terminated in finite since
$\left|Z_n\right|$ vanishes generally
when n is large. Therefore, Eq. (\ref{eq.10}) gives a well-defined equation for a
finite set of variables corresponding to $\left|Z_n\right|>0$. Putting
the solution of the equation to Eq. (\ref{eq.9}), we obtain the final
expression of $P$, and hence $S$. 
\begin{figure}
 \includegraphics{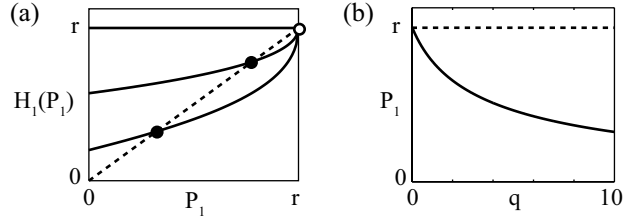}
 \caption{
 (a) (solid lines) The right hand side of the self-consistent equation
 when $q=0$, 1, and 10, from top to bottom. Dotted line is the
 left hand side of the equation. Filled and empty circles are stable and
 unstable solutions respectively. (b) Unstable (dashed line) and stable
 (thick line) solutions of the self-consistent equation.
 }
 \label{figure:1}
\end{figure}

\begin{figure}
 \includegraphics{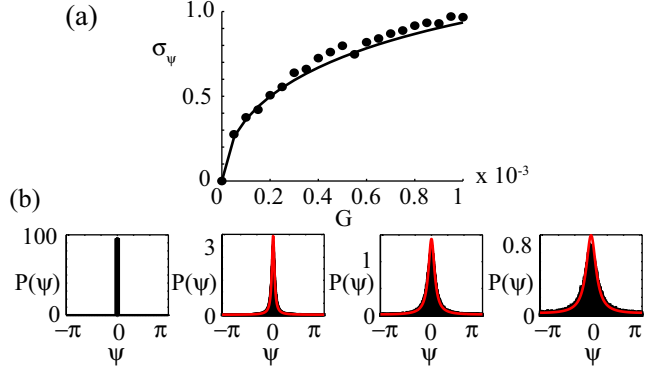}
 \caption{
 (a) Width of $P\left(\psi\right)$ as a function of the coupling strength $G$. The width is
 calculated from both theoretically (solid line) and numerically (filled
 circles). (b) $P\left(\psi\right)$ calculated numerically when $G=0.0$,
 0.2, 0.5, and 1.0 $(\times 10^{-3})$, from left to right. Solid lines (online
 red) are theoretical predictions, Eq. (\ref{eq.9}).
 }
 \label{figure:2}
\end{figure}
We now examine the above result using the simplest example where
$T=2\pi$ and $Z\left(\phi\right)=1-\cos\phi$. This phase response function is
obtained directly from the quadratic integrate-and-fire neuron, which is
also known as the theta-neuron or the Ermentrout-Kopell canonical
model\cite{ermentrout_kopell}. Putting
$\left|Z_1\right|=1/2$ and $\left|Z_n\right|=0$ for $n>1$
to Eq. (\ref{eq.10}), we obtain the self-consistent
equation for a variable $P_1$ as
$P_1=\left(1-\sqrt{1-W_1^2}\right)/(2\pi W_1)=H_1\left(P_1\right)$.
To ensure $P\geq 0$,
the condition $P_1\leq r$ is required. We plot each sides of the
self-consistent equation in Figure 1 for various values of $q$. When
$q=0$, there is no coupling in the network, the two lines have one
intersection at $P_1=r$. Inserting the solution to Eq. (\ref{eq.9}) gives
$P\left(\psi\right)=\delta\left(\psi\right)$, which exactly recover the
previous result that single neurons always response precisely and that
single oscillators always synchronize in-phase by common noise\cite{teramae}. Note
that the Fourier series of $P$ is infinite while only the 1st component
appears in the self-consistent equation. When $q>0$, i.e. there are
finite couplings in the 
network, the solution $P_1=r$ turns out to be unstable and another
stable solution appears in $P_1<r$. Inserting the solution to Eq. (\ref{eq.9}),
we obtain the distribution with a finite width. As we increase $q$,
i.e. increase coupling strength $G$, the width increases monotonically. We
plot the evolution of the width in Figure 2a. In parallel to the
analysis, we simulate Eq. (\ref{eq.2}) directly and plot the result also in
Figure 2a. Numerical results are well fitted by the theoretical
curve. Figure 2b shows distribution functions $P$ obtained from both
numerical and theoretical calculations. Theoretical predictions agree
fairly well with numerical results.

Positive width obtained here implies
that networks of oscillators, in contrast to a single oscillator, do not
synchronize perfectly across trials even though every trial is driven by
the same input. Even if all the constituents,
i.e. single oscillators here, are faithful to the input, it is not the
case for the system as a whole. Spike sequence generated by the network
is thus not perfectly the same across trials. Instead, the result
implies enough coherence of spike trains between different trials, as
indicated by the huge peak of the distribution around $\psi=0$. The
width of the peak measures a degree of the coherence, up to which spike
trains can be used reliably. Furthermore, the result tells us how the
coherence changes qualitatively as a function of the coupling strength
in the network.

If the network encodes information with precise spikes
with temporal precision $\Delta\psi$ and that the information is decoded by a downstream neuron
innervated by $K$ neurons in the network, reliability for the decoder
neuron is roughly estimated as $\left(P\left(0\right)\Delta\psi\right)^K$, which rapidly goes to
zero as $K$ increases, except if G is close to zero. Therefore the result
infers that population codes based on precise spike times require weak
or sparse mutual couplings in the network.

As the second example, we consider a coupled network of the
Hodgkin-Huxley neurons\cite{hodgkin_huxley52}. We numerically integrate the system, Eq. (\ref{eq.1}),
and compare the results predicted by theory. Since the
Fourier component $Z_n$ of the neuron decays rapidly to zero when $n>5$,
we use only the first 5 components to solve Eq. (\ref{eq.10}).
To realize the Stratonovich situation in the numerical
simulation, we use the Ornstein-Uhlenbech process with a correlation
time $\tau=1$, which is sufficiently small comparing to the period of
the oscillation. Figure 3 shows the result. As similar to the previous
example, the distribution starts from the delta-function when $G=0$ and
grows to a distribution with a finite width. The theory agrees well with numerical results,
especially when the width of the distribution is small.
\begin{figure}
 \includegraphics{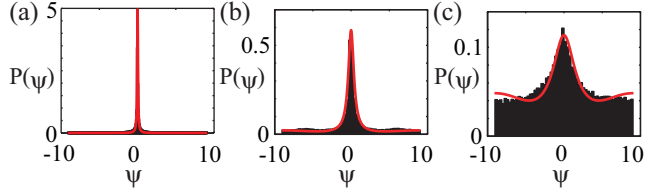}
 \caption{
 Distribution $P\left(\psi\right)$ of Hodgkin-Huxley neuron with mean
 current strength $I=6.5~mA/cm^2$.
 calculated numerically from Eq. (\ref{eq.1}) for $G=10^{-3}$ (a), $10^{-2}$ (b), and $10^{-1}$
 (c). Solid lines (online red) are theoretical predictions.
 }
 \label{figure:3}
\end{figure}

We have assumed the balanced couplings, i.e.,
$\left<g_{ij}\right>=0$. If the couplings are not balanced, coefficient
$A_2$ acquire additional terms, which in turn modify the self-consistent
equation. Such biased couplings may significantly change the
intra-network synchronization property of oscillators \cite{gerstner96},
which presumably influences the spike precision or the synchronization
across trials. It is intriguing to extend the present analysis to the
case with $\left<g_{ij}\right>\neq0$. 

Since the self-consistent equation has a unique stable solution, the
obtained distribution is always realized over repeated trials even if
initial phase differences between trials are infinitesimally small. This
property reminds us of chaos theory in dynamical systems, in which a
small difference between orbits increases rapidly until it roughly
converges to a finite value characterized by the size of chaotic
attractors. Recent studies suggested that neuronal circuits can be
chaotic in a manner useful for
computations\cite{liquid,vreeswijk,battaglia07}. It remains fascinating
whether our results reveal an active role of chaos in the brain.

We thank Y. Tsubo for fruitful discussions, H. Wang for careful reading
of the manuscript, and Y. Kuramoto for helpful comments. The present
work was supported by Kakenhi (B) 50384722 and 17022036.

\end{document}